\documentstyle[12pt,epsf,cite]{article}
\newlength{\overeqskip}
\newlength{\undereqskip}
\newcommand{\nc}{\newcommand}
\nc{\eqs}[2]{\mbox{Eqs.~(\ref{#1},\,\ref{#2})}}
\nc{\eq}[1]{\mbox{Eq.~(\ref{#1})}}
\nc{\ro}[1]{{\rm #1}}
\nc{\nn}{\nonumber}

\nc{\be}{\begin{equation}}
\nc{\ee}{\end{equation}}
\nc{\bea}{\begin{eqnarray}}
\nc{\eea}{\end{eqnarray}}
\nc{\Section}[2]{\section{#2}\label{#1}}
\nc{\Bibitem}[1]{\bibitem{#1}}
\nc{\Label}[1]{\label{#1}}

\nc{\ra}{\rightarrow}
\nc{\lra}{\leftrightarrow}

\nc{\eg}{{\em e.g.} }
\nc{\etal}{{\em et.al.}}
\nc{\sgL}{{\it sign}($L$)}

\nc{\half}{{\frac{1}{2}}}
\nc{\dt}{\frac{d}{dt}}
\nc{\neut}[1]{{\nu_{#1}}}
\nc{\aneut}[1]{{\bar{\nu}_{#1}}}

\nc{\herm}{^{\dagger}}
\nc{\pri} {^{\prime}}
\nc{\ev} {\; \mbox{eV}}
\nc{\mev}{\; \mbox{MeV}}
\nc{\pp} [1]{{P_{#1}^+}}
\nc{\pmm}[1]{{P_{#1}^-}}
\nc{\mD}{{\cal D}}

\def\lsim{\;\raise0.3ex\hbox{$<$\kern-0.75em \raise-1.1ex\hbox{$\sim$}}\;}
\def\gsim{\;\raise0.3ex\hbox{$>$\kern-0.75em \raise-1.1ex\hbox{$\sim$}}\;}

\nc{\bi}{{\bibitem}}

\def\VEV#1{{\langle #1 \rangle}}

\begin{document}
%
%
\begin{titlepage}
\pagestyle{empty}
\baselineskip=21pt
\rightline{HIP-2000-54/TH}
\rightline{NORDITA 2000/92 HE}
\rightline{December 21, 2000}
\vskip .4in

\begin{center} {\Large{\bf Creation of large spatial fluctuations in \\
                             neutrino asymmetry by neutrino oscillations}}

\end{center}
\vskip .25in

\begin{center}

Kari Enqvist$^{1,2}$, Kimmo Kainulainen$^3$ and Antti Sorri$^1$\\

\vskip .2in

$^1${\it  Physics Department, University of Helsinki, \\
              P.O.\ Box 9, FIN-00014 University of Helsinki }\\
\vskip .1in
$^2${\it  Helsinki Institute of Physics\\
              P.O.\ Box 9, FIN-00014 University of Helsinki }\\
\vskip .1in
$^3${\it  NORDITA, Blegdamsvej 17, DK-2100 Copenhagen \O, Denmark}\\

\end{center}

\vskip 0.25in

\centerline{ {\bf Abstract} }
\baselineskip=18pt
\vskip 0.7truecm\noindent
We consider active-sterile neutrino oscillations in the early
universe in an inhomogeneous isocurvature background. We show that
very small initial baryonic seed-inhomogeneities can trigger a
growth of very large amplitude spatial fluctuations in lepton
asymmetry. Domains of varying asymmetry are observed to persist
for a long time despite dissipative effects. The space dependent
asymmetry profiles give rise to MSW-resonances within the domain
boundaries, enhancing dramatically the equilibration of the sterile
neutrino species. According to our one-dimensional toy-model, the
effect is so strong that almost the entire parameter space where
exponential growth of asymmetry occurs would be ruled out by
nucleosynthesis.
\\

\end{titlepage}

\baselineskip=20pt

\section{Introduction}
The present observational situation in neutrino physics appears to
favour a conventional explanation of both solar neutrino problem
and the atmospheric neutrino problem with large
angle mixing among the ordinary, known neutrino species \cite{empsit}.
However, these scenarios leave unexplained  the anomaly
observed in Los Alamos \cite{LosAl} neutrino experiment. To account for all
known anomalies one necessarily must invoke new, sterile neutrino states $\nu_s$,
mixing with ordinary neutrinos. Such states  also occur naturally in
many particle physics theories beyond the minimal standard model.
Active-sterile mixing would have several very interesting effects in
astrophysical settings \cite{snova} and in particular for the evolution
of the early universe
\cite{dkdb,ekmL,ektBig,jim,ektat,ssf,fvSc,shi,ftv,fvA,fvD,shifuNc,fvL,CMB}. 
For example, the otherwise inert sterile states would interact with ordinary
matter through mixing induced oscillation effects, and under the most
natural assumptions could, for a wide range of oscillation parameters,
be brought into thermal equilibrium prior to nucleosynthesis. The ensuing
increase in the expansion rate of the universe could then bring the
nucleosynthesis predictions for light element abundances in conflict
with the observations  \cite{dkdb,ekmL,ektBig,jim,ektat,ssf}, in
particular by causing an overabundance of helium-4. This would in
particular be the case for the active-sterile neutrino oscillation
solution for the atmospheric neutrino problem \cite{ektat}.

Even outside the region of mixing parameters where equilibration is
effective, one encounters other interesting phenomena. For negative
mass squared difference $\delta m^2 <0$, the lepton asymmetry $L$ has
been shown to evolve into an instability, which triggers an exponential
growth of $L$ at the resonance temperature $T_{\rm res}$ \cite{dkdb}. This
idea has been used to circumvent the abovementioned exclusion of the
$\nu_\mu-\nu_s$ mixing solution for the atmospheric anomaly
\cite{fvSc}.
The idea is that prior to temperature at which $\nu_s$ would be brought
into equilibrium by $\nu_\mu-\nu_s$ oscillations, a resonance in another
mixing sector, say $\nu_\tau-\nu_s'$, creates a very large lepton asymmetry
(without equilibrating $\nu_s'$), which then suppresses the amplitude
of oscillations in the $\nu_\mu-\nu_s$-sector enough to keep $\nu_s$ out
of equilibrium until after nucleosynthesis.

Later it was shown that the asymmetry growth after the resonance is 
chaotic in the sense that the final sign of the asymmetry cannot be 
simply deduced from initial conditions \cite{shi}. It was recently verified
\cite{eks1,Sorri} that this effect is real, and that the sign is sensitive
not only to the initial conditions, but for a wide region of oscillation
parameters 
(``chaotic"\footnote{With the word chaotic we do not
mean chaoticity in the mathematically strict sense, but that
the solutions are highly sensitive to both initial and boundary 
conditions.} 
region) is also is unstable against small changes in
parameters 
themselves\footnote{There has been some
debate as to whether the sign is chaotic or not. The analysis of ref.\
\cite{dolhan} has recently been shown to be flawed \cite{Sorri,DiBetal} 
while the most recent numerical studies a momentum dependent quantum 
kinetic equations are not in conflict with the results of \cite{eks1}.}.
Large asymmetries in electronic sector can significantly alter 
helium-4 abundance, which then cannot be reliably predicted should 
oscillation parameters ever be observed to lie in the chaotic region 
\cite{eks1}.

In this letter we further study the active-sterile neutrino oscillations
and exponential growth of the asymmetry.  Here we induce the additional,
and completely natural, ingredient of spatially inhomogeneous initial
conditions.  Indeed, many baryogenesis mechanisms as well as possible
first order QCD phase transition would predict isocurvature fluctuations
in the baryon asymmetry in the early universe which persist all the way
to the nucleosynthesis. What makes these fluctuations interesting for us
is that they can provide seeds for a large inhomogeneous growth of lepton
asymmetry, given the resonance conditions and the
sensitivity of the direction of growth on initial conditions. We will
show that this is indeed the case and that even very small baryonic
seed-inhomogeneities may trigger a growth of very large amplitude
fluctuations in lepton asymmetry. It should be noted however, that the
occurrence of large amplitude fluctuations in $L$ is not confined to the
chaotic region of parameter space.  We observe domain creation also in
simulations involving ``stable" mixing parameters, where the final sign
is robust and defined by the sign of the effective asymmetry at $T_{\rm
res}$. The domain structure in this case is triggered by the combination
of inhomogeneous baryon asymmetry in the background and the diffusion
effects, which together cause the effective total asymmetry seen by
neutrinos to fluctuate in sign at the time of the resonance. This
mechanism was first discussed in \cite{DiBari}.

The domains of varying asymmetry are seen to persist for a long time, 
the semistability being caused by a strong local MSW-effect which tends
to restore large $L$ against the smoothing effect of diffusion. Moreover,
the MSW-resonances within the domain boundaries lead to a rapid equilibration
of the sterile neutrinos. In our one-dimensional model this effect is so
strong that the sterile neutrinos are brought close to equilibrium over
almost the entire region of mixing parameters where the instability leading
to the $L$-growth has been observed \cite{ftv,fvA}. Would this results hold
also in the realistic 3-dimensional case, it would clearly invalidate
the scenario of avoiding $\nu_s-\nu_\mu$ equilibration, put forward in
refs.\ \cite{fvSc}, and discussed above. In short, the remedy would not
work because in the first resonance involving $\nu_{s'}$, along with
building up the asymmetry needed to prevent $\nu_s$-equilibration, the
species $\nu_{s'}$ itself would be equilibrated. However, in exchange,
one would obtain new, much stronger BBN-constraints on mixing parameters
than one finds in the spatially homogeneous computation \cite{ektBig}.

We cannot however draw firm conclusions based on our 1-dimensional calculations.  
In fact, diffusion can be expected to be more efficient in
higher dimensions, and also the realistic momentum dependent MSW-effect
could be somewhat weaker in upholding the domains. However, it is clear
that without a very careful account of the effects of inhomogeneities,
the BBN constraints on mixing parameters may be underestimated, and
moreover, the feasibility of the mechanism of \cite{fvSc} remains in
doubt.

In section 2 we will set up our formalism by a variant of a moment
expansion for the quantum kinetic equations valid for slowly varying
semiclassical background. The outcome is a slight generalization of the
single momentum approximation for QKE's with the inclusion of diffusion
corrections. In section 3 we detail our numerical approach, solve the
evolution equations and present the results in our 1-dimensional model.
Section 4 contains our conclusions and outlook.

\section{Diffusion equations}

Our starting point is the generalization of the usual Boltzmann equation
for the particle distribution function to the case of a density matrix of
a mixing 2-component system \cite{JKPTalk}:
\be
\partial_t \rho_{ij}
         + \half \{\partial_{\bf p} H,\partial_{\bf x} \rho \}_{ij}
         - \half \{\partial_{\bf x} H,\partial_{\bf p} \rho \}_{ij}
         + i [H,\rho]_{ij} = C_{ij}[\rho ].
\label{fullQBE}
\ee
The Hamiltonian $H$ of the system (we have dropped an irrelevant constant 
diagonal piece) is in the flavour basis given by
\be
H = \frac{\delta m^2}{2p} U_{\theta}\sigma_3 U^\dagger_{\theta}
   + V_\alpha \sigma^+,
\label{hamilton}
\ee
where $\delta m^2 \equiv m^2_{\neut{2}} - m^2_{\neut{1}}$,
$\sigma^+ = (1+\sigma_3)/2$ and $\theta$ 
is the vacuum mixing angle. $U_\theta$ is the usual 2x2-vacuum mixing matrix 
with the convention that the mass eigenstate $\nu_1$ becomes the active
state in the limit $\theta \rightarrow 0$ and $V_\alpha$ is the matter 
induced effective potential for the active species.  The curly brackets in 
(\ref{fullQBE}) represent anticommutators. These terms are a generalization 
of the flow derivatives of a scalar Boltzmann equation to a case of a mixing 
multicomponent system, guided by the principle of hermiticity. The commutator 
term completes the Liouville operator on the l.h.s.  The collision term 
$C_{ij}[\rho ]$ has the components (no sum over repeated indices is 
implied) 
\be
  C_{ij}[\rho ] = - D (1-\delta_{ij}) \rho_{ij} 
                  - \Gamma_{\rm el} \sigma^+_{ij} (\rho_{ij} - \rho_{\rm eq})
                  + C_{ij}^{\rm inel}[\rho ],
\label{collterm1}
\ee
where we wrote the elastic collision term in the relaxation time 
approximation and separated from it the purely off-diagonal quantum 
damping matrix with $D = \frac{1}{2} \Gamma_{el}$. The function 
$\rho_{\rm eq}(p)$ is proportional to the equilibrium distribution 
for massless fermions, with the property that the integral over the elastic 
interaction term vanishes. Finally $C^{\rm inel}[\rho ]$ is the standard 
inelastic collision integral, and it is also proportional to $\sigma^+$.

The set of {\em quantum Boltzmann equations} (\ref{fullQBE}-\ref{collterm1})  
provide an adequate description for a coupled mixing system of neutrinos in 
the presence of decohering scatterings and varying background, as long as the 
the length scale of the background variation $\ell_b$ is much larger than 
the compton wave length $\ell_C$ of particles. Here this condition is 
satisfied by an ample margin, since $\ell_b$ is some fraction of the 
Hubble radius whereas $l_C  \sim 1/p \sim 1/T$.  
 
We now make several simplifications. First, we observe that in the 
relativistic limit $H \sim p$, so that to a very good approximation 
\eq{fullQBE} reduces to
\be
\partial_t \rho_{ij} + {\bf \hat p} \cdot \partial_{\bf x}
    \rho_{ij} + i [H,\rho]_{ij} = C_{ij}[\rho],
\label{simpleQBE}
\ee
where $\hat {\bf p} \equiv {\bf p}/p$.  This equation is still a very 
complicated to solve, and we will proceed by reducing it to a truncated
set of moment equations. To this end we make the following 
decomposition
\be
\rho(p) = \tilde \rho f_{\rm eq}(p) + \delta \rho(p) ; \qquad
\int_p \delta \rho(p) = 0,
\label{ansaz}
\ee
where $\tilde \rho \equiv \rho(\VEV{p})$ and the integral condition follows 
from a normalization assumption $\int_p \rho = \tilde \rho N_{\rm eq}$. 
Integrating \eq{simpleQBE} over momentum, weighted with zeroth and first 
power of velocity (here we have simply ${\bf v} = \hat {\bf p} \equiv 
{\bf p}/p$), gives the following two equations
\bea 
\partial_t \tilde \rho_{ij} + \partial_{\bf x} \cdot {\bf \Delta}_{ij}
             + i[\tilde H, \tilde \rho]_{ij}
              &=& \langle C_{ij} \rangle 
\label{moments1}
\\  
\partial_t {\bf \Delta}_{ij}
              + \partial_{\bf x} \tilde \rho_{ij}
              + i[\tilde H, {\bf \Delta}]_{ij}
              &=& \langle \hat {\bf p} C_{ij} \rangle ,
\label{moments2}
\eea
where the thermal averages are defined as $\langle \cdots \rangle \equiv 
\int_p \cdots f_{\rm eq}(p)/N_{\rm eq}$, and we used the shorthand notations 
for the averaged Hamiltonian $\tilde H \equiv \langle H \rangle$ and the 
displacement vector $\bf \Delta \equiv \VEV{ \hat {\bf p} \delta \rho}$.
We also made the usual factorization approximation  $\VEV{[H,\hat {\bf p} 
\delta \rho]} \simeq [\tilde H,{\bf \Delta}]$.  It is straightforward to
see that the moments of the collision integral appearing in (\ref{moments1}-
\ref{moments2}) are given by 
\bea
   \langle C_{ij} \rangle 
       &=&  - (1-\delta_{ij})D \tilde \rho_{ij}
            + \sigma^+_{ij} \Gamma^{\rm inel}_{\alpha \alpha} 
                        ( n_{\rm eq}^2 - n_{\neut{\alpha}} n_{\aneut{\alpha}} ) 
\label{kolli1}
\\ 
   \langle \hat {\bf p} C_{ij} \rangle 
       &=&  - \left( \sigma^+_{ij} + \frac{1}{2}(1-\delta_{ij}) \right)
          \Gamma_{\rm el} {\bf \Delta}_{ij},
\label{kolli2}
\eea 
where $n_{\nu_\alpha} \equiv \tilde \rho_{\alpha\alpha}$ and $n_{\rm eq}$
is the equilibrium number density of a species of massless fermions.
Equations (\ref{moments1}-\ref{kolli2}) form a closed set of 12 equations 
(24 including antiparticles), which are already much more feasible for 
numerical solution than the initial full QBE's. 
Solving them, one would obtain $\tilde \rho_{ij}(t)$, which represents the 
evolution of the total number of neutrinos, together with 
${\bf \Delta}_{ij}(t)$, which describes the total spatial flow of neutrinos. 
These equations are, however, still quite
demanding to solve numerically. We shall therefore make a more constraining,
{\em diffusive} assumption, according to which the Liouville operator
acting on ${\bf \Delta}$ in \eq{moments2} can be neglected
in comparison with the collision term. We then have
\be
\partial_{\bf x} \cdot {\bf \Delta}_{ij}
          \simeq - \frac{1}{\tilde \Gamma_{ij}} 
                   \partial_{\bf x}^2 \tilde \rho_{ij}~,
\label{secondm2}
\ee
where we have defined $\tilde \Gamma_{ij} \equiv (\sigma^+_{ij} + 
\frac{1}{2}(1-\delta_{ij}) )\Gamma_{\rm eq}$. Using (\ref{secondm2}) 
we can eliminate ${\bf \Delta}$ from \eq{moments2}, yielding a single scalar 
(diffusion) equation:
\be
\partial_t \tilde \rho_{ij} + i [\tilde H, \tilde \rho]_{ij}
              + (1-\delta_{ij}) D \tilde \rho_{ij}
             =   \frac{1}{\tilde \Gamma_{ij}} 
                       \partial_{\bf x}^2 \tilde \rho_{ij} 
               + \sigma^+_{ij} {\cal C}_{\alpha\alpha},
\label{diffeqn} 
\ee
were we defined yet another shorthand notation ${\cal C}_{\alpha\alpha} 
\equiv \Gamma_{\alpha \alpha} ( n_{\rm eq}^2 - n_{\neut{\alpha}} 
n_{\aneut{\alpha}} )$.
Apart from the diffusion term $\sim \partial_{\bf x}^2 \tilde \rho_{ij}$, 
\eq{diffeqn} corresponds to the usual set of ``one-momentum" evolution 
equations for the reduced density matrix \cite{ektBig}. The diffusion term 
describes the smearing of the local fluctuations, generated by the resonant 
amplification of the asymmetry, due to the random motion of neutrinos.

It should be noted that the diffusion approximation, which is very
well satisfied in normal systems, such as mixing fluids for example,  
actually breaks down for sterile neutrinos whose interaction rate 
$\Gamma_{ss}^{\rm el}$ is very small (if not zero as we have assumed 
so far). Nevertheless, isotropically generated disturbances average 
out even in a strictly collisionless case due to free streaming (a 
feature clearly missing in the case of mixing fluids). We will use 
a small but nonzero value for $\Gamma_{ss}^{\rm el}$ to regulate our 
expressions\footnote{
It should be clear that this regulator leads to very little change in
the sterile neutrino production rate; we of course do {\em not} 
introduce any {\em inelastic} interactions for the sterile states.
}
(in practice we will take $\Gamma_{ss}^{\rm el}$ to be 
about a tenth of $\Gamma_{\alpha\alpha}^{\rm el}$). This approach 
mimics the effect of free streaming, except that it smooths out 
random noise that would be present in a realistic case. Moreover, an 
effective diffusion length is effected on the sterile species even 
when $\tilde \Gamma_{ss}=0$ by the combined effect of scattering 
in the active sector and the mixing induced by the Liouville term in 
(\ref{moments2}). This is presumably the reason why our results based on 
(\ref{diffeqn}) do not depend strongly on the actual (sufficiently 
small) value chosen for $\tilde \Gamma_{ss}$.
\vskip0.3cm

It is convenient to parameterize the reduced density matrices of the
neutrino and anti-neutrino ensembles in terms of the Bloch-vector
representation:
\be
\tilde \rho_{\nu} ({\bf x}) \equiv \half  \left( P_0 ({\bf x})
                      + {\bf  P}({\bf x})\cdot {\bf \vec{\sigma}} \right)
                      \; , \ \qquad
\tilde \rho_{\bar \nu} ({\bf x}) \equiv \half \left(  \bar P_0 ({\bf x})
                      + {\bf \bar P}({\bf x})\cdot {\bf \vec{\sigma}} \right),
\label{rho} 
\ee
where $\vec{\sigma}$ are the Pauli matrices. The coupled equations with the
diffusion terms from \eq{diffeqn} in the case of $\neut{\tau} - \neut{s}$
oscillations then read (other cases can easily be obtained by a simple
redefinition \cite{ektBig}):
\def\phm{\phantom{-}}
\bea
\dot {P_0}
       &=&   \mD_+ \partial_{\bf x}^2 P_0
           + \mD_- \partial_{\bf x}^2 P_z 
           + {\cal C}_{\tau\tau} \\ \nn
\dot {\bar{P}_0}
       &=&   \mD_+ \partial_{\bf x}^2 \bar P_0
           + \mD_- \partial_{\bf x}^2 \bar P_z 
           + \bar {\cal C}_{\tau\tau} \\ \nn
\dot {\bf P}
       &=& {\bf V}\times{\bf P}- D {\bf P}_T  \\ \nn
       &+& \left[ \mD_- \partial_{\bf x}^2 P_0
                + \mD_+ \partial_{\bf x}^2 P_z \right.
         \left. + {\cal C}_{\tau\tau} \right] {\bf \hat z}
           + \mD_{\tau s} \partial_{\bf x}^2 {\bf P}_T  \\ \nn
\dot {\bf {\bar P}}
       &=& {\bf \bar V}\times{\bf \bar P} - \bar D {\bf \bar P}_T \\ \nn
       &+& \left[ \mD_- \partial_{\bf x}^2 \bar P_0
                + \mD_+ \partial_{\bf x}^2 \bar P_z \right.
         \left. + \bar {\cal C}_{\tau\tau} \right] {\bf \hat z}
                + \mD_{\tau s} \partial_{\bf x}^2 {\bf \bar P}_T,
\label{one state}
\eea
where $\dot x \equiv dx/dt$, ${\bf P}_T = P_x {\bf \hat x} + P_y {\bf \hat y}$
and $\bar {\cal C}_{\tau\tau} \equiv  \bar \Gamma_{\tau \tau} 
( n_{\rm eq}^2 - n_{\neut{\tau}} n_{\aneut{\tau}} )$. The diffusion 
coefficients 
\bea
    \mD_{\pm}  &=& \frac{1}{2}\left( \frac{1}{\Gamma_{\tau \tau}^{\rm el}}
                   \pm\frac{1}{\Gamma_{s s}^{\rm el}}\right) , \\ \nn
    \mD_{\tau s}   &=& \frac{2}{\Gamma_{\tau \tau}^{\rm el} 
                       + \Gamma_{s s}^{\rm el}}
\label{difconst} 
\eea
should not be confused with damping coefficients $D$ and $\bar D$. 
The numerical values of the rates appearing above are
\cite{gio,ektBig} $\Gamma_{\tau\tau} = \bar \Gamma_{\tau\tau} 
\simeq 0.32 G_F^2T^5$ and $\bar D \simeq D  = 
\frac{1}{2}\Gamma_{\tau\tau }^{\rm el} \simeq 1.4 G_F^2 T^5$.
The rotation  vector ${\bf V}$ in the particle sector is
\be
     {\bf V} = V_x \;{\bf \hat{x}} + \bigl( V_0 + V_L \bigr) \; {\bf \hat{z}},
\label{rvec} 
\ee
with the components  
\bea
     V_x    &=&  \frac{\delta m^2}{2 \langle p \rangle} \sin 2 \theta  
\\ 
     V_{0}  &=& -\frac{\delta m^2}{2\langle p \rangle } \cos 2 \theta 
                - 17.8  G_F N_\gamma \frac{\langle p \rangle T}{2 M_Z^2} 
\\
     V_L    &=&  \sqrt{2} G_F N_{\gamma} \; L.
\label{Vcomp} 
\eea
Here $N_{\gamma}$ is the photon number density and the effective asymmetry 
$L$ in the potential $V_L$ is given by
\be
      L = - \half L_n + L_{\neut{e}} + L_{\neut{\mu}} + 2 L_{\neut{\tau}}(P)
        \equiv L_0 +  2 L_{\neut{\tau}}(P)
\label{alla} 
\ee
in the case of an electrically neutral plasma. Here $L_n$ is the neutron
asymmetry and the term $L_{\neut{\tau}}$ introduces the coupling between
the particle and antiparticle sectors:
\be
     L_{\neut{\tau}} = \frac{3}{8} \left[ \half (P_0+P_z)
                       - \half (\bar P_0 + \bar P_z ) \right].
\label{asymm} 
\ee
Finally, the rotation vector for anti-neutrinos is obtained by simply
changing the sign of the asymmetry: ${\bf \bar V}(L) = {\bf V}(-L)$.

The neutrino and anti-neutrino ensembles are very strongly coupled in
\eq{one state} through the effective potential term $V_L(L)$. This makes
their numerical solution very difficult. In particular, a technical problem
is posed by the large cancellation in \eq{asymm}, where two ${\cal O}
(1)$-terms cancel up to 10 decimal places to produce a number of order
${\cal O}(L_0)$ initially, with only slight improvement of matters before
the very final stages of calculation when $L$ becomes very large. This
obviously gives rise to a very dangerous loss of accuracy in the numerical
solution, which however can be easily avoided by defining new ``large" and
``small" variables:
\be
    P_{\mu}^\pm \equiv P_{\mu} \pm \bar{P}_{\mu}.
\label{pmvar} 
\ee
With these definitions the evolution equations become
\bea
\dot P_0^+  &=&   \phm  \mD_+\partial^2_{{\bf x}} P_0^+
                    + \mD_-\partial^2_{{\bf x}} P_z^+
                    + 2{\cal C_{\tau\tau}} \\ \nn
\dot P_0^-  &=&    \phm \mD_+ \partial^2_{\bf x} P_0^-
                    + \mD_- \partial^2_{\bf x} P_z^-  \\ \nn
\dot{P}_z^+ &=&    \phm V_x P_y^+
                     + \mD_-\partial^2_{{\bf x}} P_0^+
                     + \mD_+\partial^2_{{\bf x}} P_z^+
                    + 2{\cal C_{\tau\tau}} \\ \nn
\dot{P}_z^- &=&   \phm  V_x P_y^-
                     + \mD_-\partial^2_{{\bf x}} P_0^-
                     + \mD_+\partial^2_{{\bf x}} P_z^-  \\ \nn
\dot{P}_x^{\pm}  &=&  -   V_0 P_y^{\pm} - V_L P_y^{\mp} - D P_x^{\pm}
                       + \mD_{\tau s} \partial_{\bf x}^2 P_x^{\pm} \\ \nn
\dot{P}_y^{\pm}  &=& \phm V_0 P_x^{\pm} + V_L  P_x^{\mp}- D P_y^{\pm}
                                    -V_x P_z^{\pm}
                                    + \mD_{\tau s}
                                      \partial_{\bf x}^2 P_y^{\pm},
\label{mastereqs} 
\eea
Note in particular that the
difference $P_0^-$ is not affected by collisions. This occurs because in the
momentum averaged approximation $P_0$ is proportional
to the total number of particles which can only be changed by annihilations,
which affect particle and antiparticle systems in exactly the same way
($\Gamma_{\alpha \alpha} = \bar \Gamma_{\bar \alpha \alpha }$).

\vskip 0.2truecm

It is now well known that in the homogeneous case the \eq{mastereqs} can
lead to a period of exponential growth of asymmetry \cite{fvA,ftv,dkdb} after
the instability sets in at the resonant temperature \cite{ektBig}
\be
T_{\rm res} \simeq 16.0 \; (|\delta m^2| \cos 2\theta)^{1/6} \; \mev.
\label{Tres} 
\ee
This temperature sets the overall scale of importance for neutrino-oscillations
in the early universe. The exponential growth of $L$ is the key factor also
for our present study, because
it can lead to amplification of the spatial variations in leptonic asymmetry,
seeded by the fluctuations in baryonic asymmetry contained in the term $L_0$
in \eq{alla}. These fluctuations are the most crucial feature of our model,
and they naturally arise via several mechanism in yet earlier stages of the
evolution of the universe, such as electroweak and QCD phase transitions
\cite{kkss}.

To be concrete, we assume that the baryonic domain size is roughly one hundredth of
the Hubble radius at the time of the QCD phase transition, $d_B\simeq 0.01/H(t_{\rm
QCD})$. Inserting the value of the Hubble expansion rate $H(t_{\rm QCD})$ and
accounting for the expansion of the universe, it then follows that at the
initial temperature $T_0<m_\mu$ of our calculation the size of each distinct
lepton domain is roughly given by
\be
d_L\simeq 0.0018{M_{Pl}\over T_{\rm QCD}T}~.
\label{domainsize} 
\ee
Moreover, we have we have assumed that in each domain $L_0 (x)$ has a fixed
value, which varies randomly in the range $[0.5 \times 10^{-10}, 10^{-10}]$.
This is very conservative assumption, and fluctuations with orders of magnitude
larger amplitudes can easily be created during the QCD transition \cite{kkss}.
The domains created in electroweak transition are typically smaller by a
factor of thousand in size, and hence too small for us to resolve with our
present numerical methods. It should be stressed that these parameters are
just particular physically motivated choices; neither the actual domain
size or the range of amplitude variation have particular importance for our
final results. All that matters is that {\it some} fluctuations exist to
launch the instability.

Solving equations (\ref{mastereqs}) is numerically relatively easy in
comparison with the full momentum dependent equations, resulting in huge
drop in the computer power required. We believe that they nevertheless
provide a qualitatively correct description of the essential features of asymmetry
oscillations in the spatially varying background. In fact, the numerical
difficulties inherent in the full momentum dependent equations are so severe
that at present the asymmetry oscillations can not be distinguished with
certainty from computer-generated numerical errors even in the spatially
homogeneous case \cite{fdb}, and hence using them in the spatially varying
case would appear an insurmountable task.

Let us finally comment on feasibility for the present problem of the
so-called static approximation \cite{fvA,ftv}, which is often used to simplify
the full momentum dependent quantum kinetic equations. The static
approximation does include momentum dependence, but has the great
disadvantage that it averages out the asymmetry oscillations, whereby the
effect of creation of leptonic domains in the chaotic region is poorly
described in that approximation.  More importantly, the static approximation
excludes the MSW-resonance occurring in the boundaries of the domains. Since
these resonances are responsible for the abundant production of sterile
neutrinos we are finding in our computations, we conclude that the static
approximation fails to describe the most striking feature of the asymmetry
oscillations in the inhomogeneous backgrounds. Hence the only reasonable
improvement over the present approximation is to go directly to full QKE's,
which, as stated above, appears an overly difficult task to be implemented
numerically.

The main advantage of evolution equations (\ref{mastereqs}), over the
previous treatments \cite{shifu2,DiBari} is the inclusion of the diffusion
terms while keeping a full account of the asymmetry oscillations and the
MSW-effect. This, we shall see, leads to new and interesting features in
the solutions.

\section{Numerical results}

\begin{figure}[ht]
\centering
\leavevmode\epsfysize=10cm \epsfbox{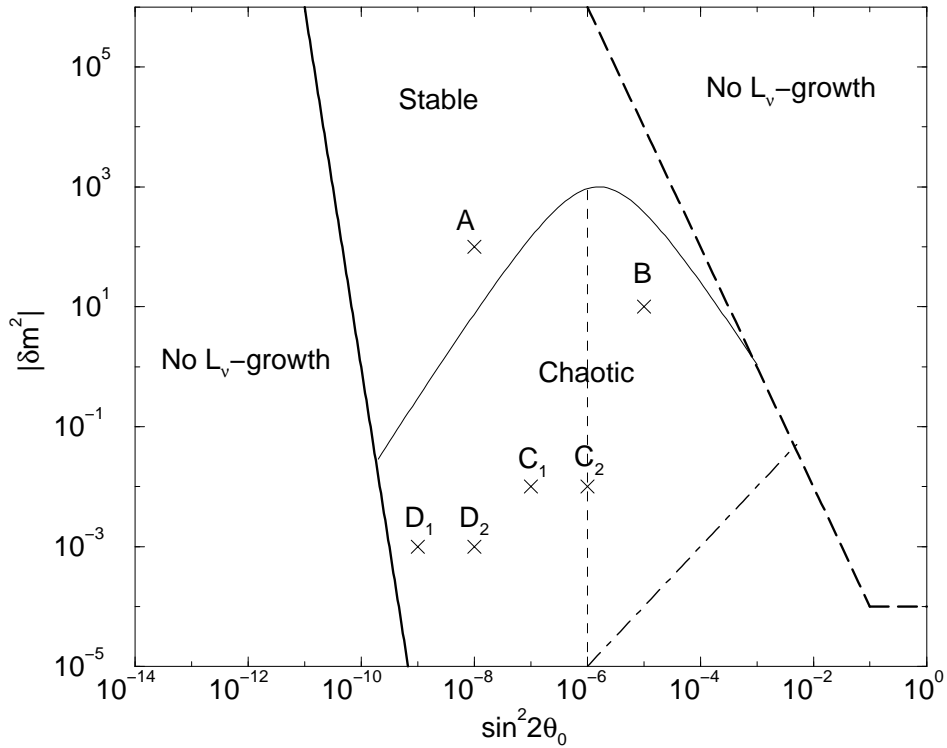}
\caption{The mixing parameter space divided into regions with qualitatively
different phenomenological aspects. }
\label{fig:plane}
\end{figure}

It is useful first to consider the information provided by earlier studies
without spatial fluctuations as a function of the mixing parameters.
Indeed, the $(\delta m^2, \sin^22\theta_0)$-plane can be divided to regions
with phenomenologically distinct characteristics as is roughly outlined in
Fig.~\ref{fig:plane}.  In the region left from the thick solid line there
is no growth of asymmetry \cite{fvA,ftv}. This occurs because for very
small mixing angles the resonance becomes so nonadiabatic that the solution
has no time to pull away from the weak local fixed point at $L=0$, which
still exists below the resonance temperature for very small mixing. Using
results of \cite{ftv} one can estimate that this occurs for
$\sin^22\theta_0 |\delta m^2|^{1/6} \lsim 10^{-10}$.
The asymmetry growth is thwarted also in the region to the right from the
thick dashed line, because there the sterile states are brought to a full
thermal equilibrium before the resonance \cite{ektBig}. A large asymmetry
growth is observed in the rest of the parameter space. This
region is further subdivided to three areas: the area above the thin solid
line is called ``stable region", since there the final sign of the neutrino
asymmetry is determined by the initial baryon asymmetry.  Below the thin
solid line lies ``chaotic region", where the final sign of the neutrino
asymmetry is very sensitive to tiny changes in either initial conditions,
or oscillation parameters, due to a period of rapid sign changing oscillations
of the neutrino asymmetry right after the resonance. Slightly different
results to the extent of the chaotic region exist in the literature:
the region bounded by the thin solid line corresponds to the results of
\cite{eks1}, whereas \cite{fdb} reports a possibility of chaotic behaviour
in a somewhat smaller area, extending to the right from the thin dashed
line. This difference is likely attributable to the fact that \cite{fdb}
used the momentum dependent QKE's, whereas a momentum averaged
equations were employed in \cite{eks1}. Finally, in the lower right
corner below the dash-dotted line the asymmetry oscillations continue
until the active neutrino decoupling. Since our diffusive approximation
is not valid anymore on this area, we shall not consider these
parameters further in this paper.

\vskip 0.3truecm
\noindent{\bf Numerical setup.}
We have solved the set of partial differential equations \eq{mastereqs} by
the method of lines using central finite differences.  Obviously the spatial discretization must be such that
our grid encompasses all the relevant physical scales. On the small distance
side this implies that the grid must be dense enough to accurately model
the initial domains of baryon asymmetry. Our numerical checks indicate that
each domain should be divided to at least twenty sub-zones, to get sufficient
accuracy in evaluation of the numerical derivatives in diffusion terms.
On the other hand, the grid must be large enough to accommodate all
diffusion scales (active and sterile) during the entire computation.
This is  particularly complicated, because the diffusion length
increases as $\ell_D \sim T^{-5}$ as the temperature decreases. These
constraints on the discretization lead to the need for very large lattices,
which require a lot of computer power. Thus, for a first study, we have only
considered one dimensional space. This is of course unphysical and to a degree
compromises our ability to draw definitive quantitative conclusions, but we 
believe that our results provide important qualitative insight to the physical
phenomenon under study.  We will return to this issue in our conclusions.

The complete solution of the problem requires setting the boundary conditions
at the ends of the lattice. We have used periodic boundary conditions, which
are very natural, since they automatically impose the desired conservation of
particle number. One might think however, that periodicity is too restrictive
when assumed also for the off-diagonals. We have hence investigated the effect
of using various other possible boundary conditions on the system (cf.\ Fig.\
3), but found no difference in results, which could not be attributed to finite
size effects, as long as the particle number was conserved.

To cope with the requirements on the grid size, introduced by the rapid
growth of $\ell_D$, we invented a technical trick of doubling the grid size
periodically. The doubling step consists of simply discarding every second
point in the original grid and replacing it with a double image of the
resulting more sparse grid.  This procedure can be repeated as many times
as necessary, provided that no relevant physics is lost when coarse graining
steps are taken.  Certainly the initial baryon number fluctuations, whose
scale remains frozen, are eventually evened out by the coarse graining.
However, as will be explained later, these features are not relevant for
the dynamics of the system  immediately after the resonance. Indeed,
we have checked the validity of the doubling method by comparing results of the
codes running with and without a doubling step over an interval of time
where both are expected to be valid, finding an excellent agreement.

\vskip 0.3truecm
We now describe the characteristic features of our numerical solutions. As
in the spatially homogeneous case, the evolution of the
asymmetry can be divided to three different phases, albeit the physics
at each stage is markedly different.

\vskip 0.3truecm
\noindent{\bf 1. Pre-resonance region.}
Above the resonance temperature (\ref{Tres}) the effective asymmetry within
each domain is driven towards a stable fixed point where $L=0$ (cf \eq{alla})
by oscillations. This behaviour is very similar to the homogeneous case, except
that the frozen-in baryonic fluctuations in the background asymmetry $L_0$ now
induce compensating spatial variations in the dynamical variable $L_{\nu_\tau}
(P({\bf x}))$.  On the other hand, diffusion tends to smooth out these
variations, thereby opposing the tendency of relaxation to $L({\bf x}) = 0$.
The outcome is that just before the resonance the effective asymmetry
$L({\bf x})$ will have spatial fluctuations, whose amplitude and scale are
set by the initial amplitude and the size of the domains and the strength
of the diffusion at the resonance. These fluctuations also typically oscillate
in sign even though the initial fluctuations were all positive, as was recently
noted by DiBari \cite{DiBari}. The sign fluctuations are requisite for the
creation of the large amplitude fluctuations in $L$ in the stable region of
parameters. In the chaotic region the asymmetry generation is not dependent
on this effect however, and there a spatially varying $L_{\nu_\tau}
(P({\bf x}))$ at the onset of the resonance is sufficient. In Fig. 2 we have
plotted the variable $L_{\nu_\tau}(P({\bf x}))$ as a function of ${\bf x}$
for a small part of our grid in various temperatures for  $\sin^22\theta_0 =
10^{-5}$ and $\delta m^2=-1 \;{\rm eV}^2$. The  uppermost panel corresponds
to a situation a little above the resonance, which for these parameters occurs
at $T_{\rm res} \simeq 16$ MeV.  The distribution in this panel is effectively
the negative of our initial baryon asymmetry fluctuation spectrum, smoothed
out by diffusion.

\vskip 0.3truecm
\noindent{\bf 2. Resonance region.}
At the resonance the lepton asymmetry enters the region of instability. In
other words, the previously stable fixed point $L=0$ becomes unstable, and
asymmetry undergoes a period of exponential growth and sometimes violent
oscillations (chaotic region).  In the stable region, the domains of opposite
effective asymmetry $L({\bf x})$ start to grow rapidly in opposite directions.
Similarly in the chaotic region, the spatially varying 
$L_{\nu_\tau}(P({\bf x)})$
triggers a period of growth and incoherent oscillation of asymmetry in
neighbouring regions. In both cases this behaviour is regulated by the
diffusion, which tries to smooth out the fluctuations just generated. The
oscillation induced asymmetry growth is much stronger however, and asymmetry
growth cannot be stopped. Second panel in Fig.\ 2 shows how the spatial
variations in $L$ looks during the rapid oscillation period following the
resonance this point. It should be noted that soon after the initial
exponential growth period the dynamics of the system is solely determined
by balancing diffusion against the MSW-effect, which is driven by {\em the
instantaneous} asymmetry configuration; the initial asymmetry $L_0$ plays
no role at all, apart from triggering the initial growth period.

\begin{figure}[ht]
\centering
\leavevmode\epsfysize=6.5cm \epsfbox{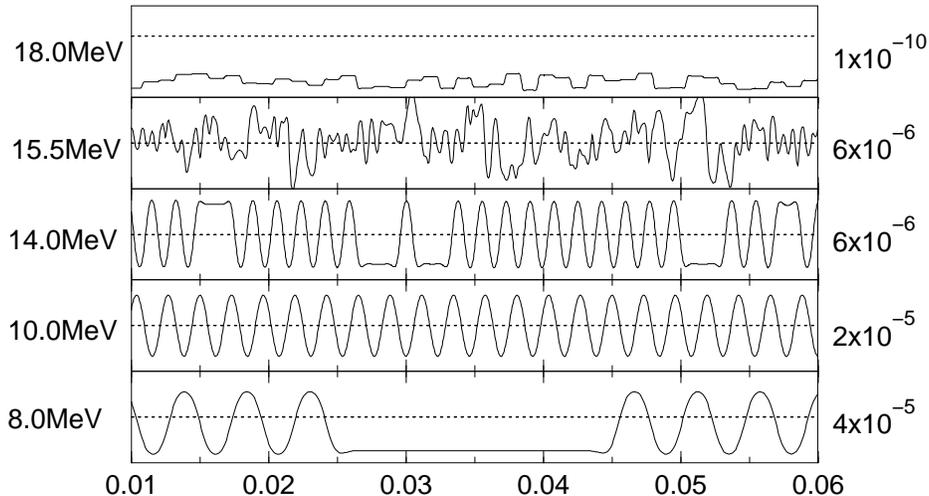}
\caption{Shown is the spatial variation of the neutrino asymmetry for five
representative temperatures, indicated by the numbers to the left 
from the panels,
for $(\sin^22\theta_0,\delta m^2) = (10^{-5},-1\; {\rm eV}^2)$. The units are
$\ell_{18}(18{\rm MeV}/T)$ where $\ell_{18}$ is the Hubble radius at 
$T=18$ MeV,
and the numbers to the right give the extent of the $y$-axis in each subpanel,
measured from bottom to the dotted line indicating $L_{\nu_\tau}=0$. Only
a small part of the actual grid is shown.}
\label{fig:evo}
\end{figure}

\vskip 0.3truecm
\noindent{\bf 3. Annihilation region.}
After the resonance, once the asymmetry oscillations end and the asymmetry
growth gets less violent in the chaotic region, diffusion effects take
partly over.  At this stage a quasistable distribution of leptonic
domains with varying sign of $L$ is created. The smallest domain size at this
point is set by the diffusion length $\ell_D$ of the active species, while a
typical domain size is a few times diffusion length and some of the largest
may be even ten times the diffusion length in size. The absolute value of
the asymmetry in the domain centers approaches the new power law fixed point
value of the homogeneous case~\cite{fvA,ftv,DiBetal,shi}:
\be
   L_{\rm fp} \propto \pm T^{-4}, \quad T<T_{\rm res},
\label{fixed} 
\ee
with $|L_{\rm fp}| \gg |L_0|$. After this point the number of leptonic
domains remains unchanged for some time, while the size of the domains
adjusts slowly.  Here one observes the peculiar phenomenon that the smallest 
domains do not vanish, but instead {\em grow} with the diffusion length 
at the expense of larger ones. This is caused by the extremely strong 
tendency of a local asymmetry to roll towards the fixed point value 
(\ref{fixed}). This ``pull" gets stronger the further away from the fixed 
point $L$ is driven, and thus, it is the relative size of the large domains 
that gives away first to diffusion rather than annihilating away the smaller 
ones. This stage of evolution is depicted in the third panel 
in Fig.~\ref{fig:evo}.  In the stable case large leptonic asymmetries are 
first created in boundaries domains with different baryonic asymmetry. 
The number and size of created domains is then initially sensitive to the 
size of the baryonic domains. However, when the diffusion effects kick in, 
the later evolution of the system is very similar to the one observed in 
the chaotic region; only the mechanism of domain generation is different.

At some point, the smaller domains have eaten the larger domains resulting
in a distribution of equal sized domains (fourth panel in Fig.\ 2), after 
which domains eventually  begin to annihilate.  However, as was explained 
above, there is a resisting tendency against domain annihilation, so that 
(in a finite grid) the system ``supercools" against  domain annihilation. As a
result, when an annihilation eventually takes place, several domains are
often destroyed at once (fifth panel in Fig.\ 2).  In a finite grid such a
multiple annihilation event shows up as a slight discontinuity in the sterile 
neutrino production rate. After an annihilation event the equalizing effect 
takes over and smooths the distribution again, until an  even distribution of 
yet larger  domains is again established, and the cycle  takes over.  It 
should be obvious that eventually  the diffusion  effect must win, 
completely smoothing out the asymmetry.   However, our approximations 
break down well before this would happen and  hence we cannot see this 
limit in  our numerical examples.

\begin{figure}[ht]
\centering
\leavevmode\epsfysize=10cm \epsfbox{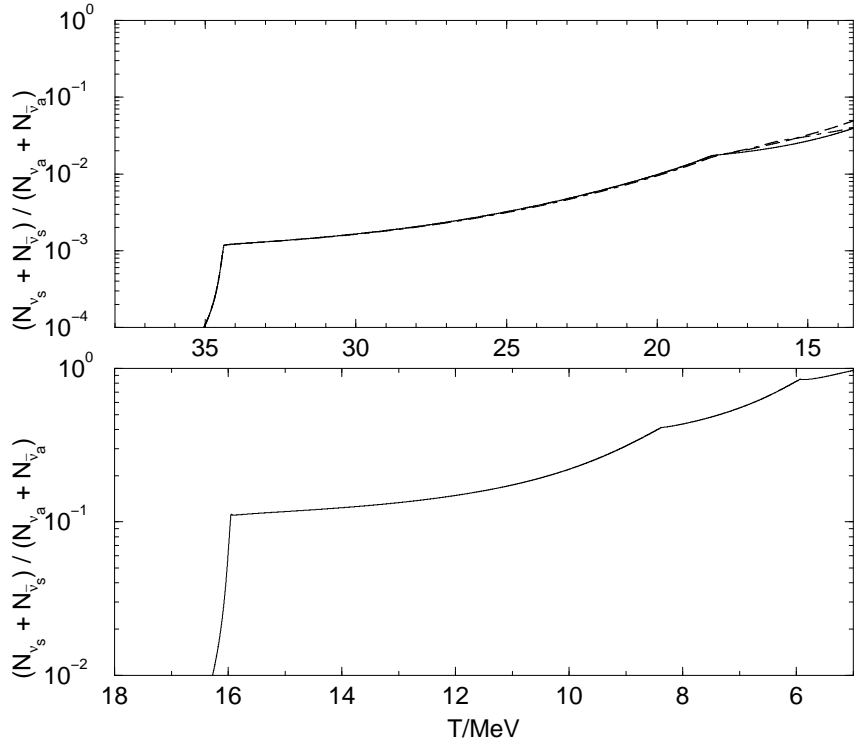}
\caption{Ratio of sterile and active neutrino number densities as a function
of temperature. Upper (lower) figure corresponds to parameters indicated by
points $A$ ($B$) in Fig.\ 1. The dashed and dash-dotted lines correspond to
set $A$ with different boundary conditions (see text).}
\label{fig:ppav}
\end{figure}

\vskip0.3truecm
\noindent{\bf Physical consequences.}
The most important physical implication of the existence of the quasistable
domains of varying asymmetry is that it gives rise to copious production of
sterile neutrinos, as a result of an MSW-effect on the domain boundaries,
induced by the spatially varying effective potential $V_L \propto L$. Indeed,
the large mixing at the resonance, combined with the decohering interactions
brings sterile neutrinos into equilibrium very efficiently. This mechanism
was first discussed heuristically by Shi and Fuller \cite{shifu2}. Here
the mechanism is automatically built into our evolution equations. In Fig.\
\ref{fig:ppav} we show the ratio of the (sum of neutrinos and antineutrinos)
active and the sterile number densities as a function of temperature for two
representative choices of parameters. The upper figure corresponds to
$(\sin^22\theta_0,\delta m^2) = (10^{-8},-10^2 \ev^2)$, taken from the
stable region of parameters (the point $A$ in Fig.\ 1 ), and the lower
figure corresponds to $(\sin^22\theta_0,\delta m^2) = (10^{-5},-1 \ev^2)$
from within the chaotic region (point $B$). The rapid initial growth observed
in each of the curves corresponds to the pre-instability near resonance when
$L\approx 0$ throughout the space.  Immediately after the resonance $L$
grows rapidly and the MSW-effect becomes confined to within narrow bands
inside the domain boundaries. Moreover, it occurs only for particles
($L>0$) or antiparticles ($L<0$) at a time. As a result the equilibration
proceeds slower after resonance. 

The dashed and dash-dotted lines in the upper figure correspond to using 
different boundary conditions.  For dashed line we imposed periodicity for 
the diagonals of $\rho$, but set the off-diagonals  to zero at the boundaries. 
For dash-dotted line we imposed periodicity for all first derivatives of $\rho$, 
but did not restrict $\rho$ itself. The  difference in results is essentially 
a measure of the finite size effects, and become nonneglible only when the
structure  of equally sized domains is settling in.  Slight ``dents" seen in 
the upper figure, caused by the multiple  domain annihilations, are another
reflection of finite size of  the grid. We  have checked that these features
become less  prominent when the  lattice size  is increased, without changing our
conclusions. We have also  checked that the  production of sterile neutrinos is
robust to variations in  the cut-off value  of sterile neutrino interaction rate.
In fact the effect  of decreasing sterile  neutrino interaction rate is a small
increase in the  rate of sterile neutrino production, which helps to justify our
diffusive  approximation.

In the one dimensional case studied here the equilibration effect is very
strong: while we have not tried to completely map the parameter space, it
appears clear that for most of the parameter space where exponential asymmetry
growth takes place, the sterile neutrinos are excited with large enough energy
density to be in conflict with nucleosynthesis constraints~\cite{t_limits}.
We have shown this explicitly in Fig.\ 3 for parameters corresponding to
points $A$ and $B$ in Fig.\ 1. However, one does expect that for very small
$|\delta m^2|$ no equilibration should take place, because then $T_{\rm res}$
becomes very low, and consequently the overall scattering rate $\Gamma \sim
T^5$ becomes very weak. We have checked that for sets $C_1$
and $C_2$ we still find relatively strong equilibration, whereas for sets
$D_1$ and $D_2$, corresponding to $T_{\rm res} \simeq 5$ MeV, practically no
$\nu_{s}$-excitation takes place. So, below a line situated somewhere between
these sets exists a region where a large $L$ {\em can} be created without
bringing $\nu_s$ into equilibrium. However, this region is already
disfavoured by other reasons \cite{fvD}. We have also
checked for several parameter sets that  for very small mixing angles (region
left from thick solid line in Fig.\ 1) no sterile neutrino equilibration takes
place.

Another interesting result following from our studies is that, despite the
large fluctuations in leptonic asymmetry, the {\rm average} total asymmetry
is found to be zero in our calculations, up to finite size effects. 
This is of course not surprising, but
rather something one should naturally expect even without a detailed
computation; an average of a large number of sub-horizon scale fluctuations
with random signs and symmetric amplitudes should be close to zero. Notably,
such a configuration has much less effect for nucleosynthesis than would
have a large scale homogeneous lepton asymmetry \cite{shifuNc}.

Should our results hold also in a realistic 3-dimensional case, BBN would
exclude the active-sterile neutrino mixing essentially in the entire
interesting region of parameters~\cite{fvD}.  However, one should
expect that both the inclusion of the realistic momentum distributions, and
the higher dimensional spatial structure could decrease the strength of our
effect. Introducing the momentum distribution would spread out the
MSW-resonance condition on the wall boundaries, perhaps making it less
efficient in opposing the diffusion effects. This might lead to somewhat
shallower boundaries and perhaps easier domain annihilation, to which we
already alluded above.  In a higher dimensional lattice the diffusion would
be much more efficient simply because of the topology; now the domains could
spread out in more ways than one, and the tendency to create a quasistable
domain configuration would be weaker. Unfortunately such simulations turn
out to be too demanding numerically at the moment, and we cannot provide
any quantitative estimate for the size of these effects. We therefore
will only conclude that we have found, and verified in a toy model, an
interesting new feature of asymmetry oscillations in the early universe,
which has the potential to significantly affect (strengthen) the bounds
on neutrino mixing parameters arising from BBN-considerations.

%
%

\section{Conclusions}

We have shown that large amplitude fluctuations in the lepton asymmetry can
naturally be created in the early Universe. The main ingredients needed are
the active-sterile neutrino mixing, which for a large range of mixing
parameters encounters an instability leading to a period of exponential
growth of asymmetry at $T_{\rm res} \sim {\cal O}(10)$ MeV, and small
baryonic seed asymmetries, which are naturally created for example during
the electroweak and QCD phase transitions. The effect is not dependent on
the particular values of amplitudes and distance scales of the seed
asymmetries, which are only needed to trigger the initial inhomogeneous
growth, because right after the initial growth period, the dynamics of 
the system
is solely governed by the MSW-effect and diffusion.

The phenomenology of the system is completely different from what one finds
in calculations assuming a homogeneous background.  First, the asymmetry does
not reach a large homogeneous value during the BBN. Instead, a quasistable
structure of sub-horizon domains with varying sign of asymmetry is realized,
while the average asymmetry over the horizon scale remains close to zero.
Such a configuration has much less impact on the light element abundances
\cite{shifuNc}, or cosmic microwave background \cite{footbari}
than a large homogeneous asymmetry would have had. Moreover,
it is important to note that the creation of sub-horizon domains is a robust
feature. In order to realize a scenario where only superhorizon scale
perturbations are amplified \cite{DiBari}, one must make the unnatural
ad hoc assumption that the early universe is almost exactly homogeneous
at all sub-horizon scales. If this were not the case, the tiny sub-horizon
scale fluctuations would take over, giving rise to the structures discussed in
this paper and  thereby completely drowning any signature at superhorizon
scales.

An MSW-resonance would occur within the domain boundaries, where 
$V_L \propto L({\bf x})$, in a manner 
very much similar to the inside of the Sun.  As
a result the mixing angle would be enhanced within a confined region inside each
domain, which, together with the rapid decohering scatterings, leads to an
efficient equilibration of the sterile species. We found
this effect to be strong enough to lead into conflict with BBN-constraints
over most of the region where the rapid asymmetry growth has been detected
in  homogeneous calculations.  However, we stress that our computations are based on
a one-dimensional toy model, and furthermore employ a momentum averaged
approximation for the full QKE's. Relaxing either of these approximations
might be expected to weaken the equilibration effect, although it is
impossible to quantify how much.  We conclude that the presence of baryonic
seed inhomogeneities may have a major impact on sterile-active
neutrino mixing in the early universe, and in particular on the BBN-bounds
on mixing parameters.

\vskip 0.5truecm

%
%

\section*{Acknowledgements}

\noindent This work has been supported by the Academy of Finland  under
the contract 101-35224. We thank the Centre for Scientific Computing
(Finland) for providing computer resources.

%
%
\nc{\advp}[3]{{\it  Adv.\ in\ Phys.\ }{{\bf #1} {(#2)} {#3}}}
\nc{\annp}[3]{{\it  Ann.\ Phys.\ (N.Y.)\ }{{\bf #1} {(#2)} {#3}}}
\nc{\apl}[3] {{\it  Appl. Phys. Lett. }{{\bf #1} {(#2)} {#3}}}
\nc{\apj}[3] {{\it  Ap.\ J.\ }{{\bf #1} {(#2)} {#3}}}
\nc{\apjl}[3]{{\it  Ap.\ J.\ Lett.\ }{{\bf #1} {(#2)} {#3}}}
\nc{\app}[3] {{\it  Astropart.\ Phys.\ }{{\bf #1} {(#2)} {#3}}}
\nc{\cmp}[3] {{\it  Comm.\ Math.\ Phys.\ }{{ \bf #1} {(#2)} {#3}}}
\nc{\cqg}[3] {{\it  Class.\ Quant.\ Grav.\ }{{\bf #1} {(#2)} {#3}}}
\nc{\epl}[3] {{\it  Europhys.\ Lett.\ }{{\bf #1} {(#2)} {#3}}}
\nc{\ijmp}[3]{{\it  Int.\ J.\ Mod.\ Phys.\ }{{\bf #1} {(#2)} {#3}}}
\nc{\ijtp}[3]{{\it  Int.\ J.\ Theor.\ Phys.\ }{{\bf #1} {(#2)} {#3}}}
\nc{\jmp}[3] {{\it  J.\ Math.\ Phys.\ }{{ \bf #1} {(#2)} {#3}}}
\nc{\jpa}[3] {{\it  J.\ Phys.\ A\ }{{\bf #1} {(#2)} {#3}}}
\nc{\jpc}[3] {{\it  J.\ Phys.\ C\ }{{\bf #1} {(#2)} {#3}}}
\nc{\jap}[3] {{\it  J.\ Appl.\ Phys.\ }{{\bf #1} {(#2)} {#3}}}
\nc{\jpsj}[3]{{\it  J.\ Phys.\ Soc.\ Japan\ }{{\bf #1} {(#2)} {#3}}}
\nc{\lmp}[3] {{\it  Lett.\ Math.\ Phys.\ }{{\bf #1} {(#2)} {#3}}}
\nc{\mpl}[3] {{\it  Mod.\ Phys.\ Lett.\ }{{\bf #1} {(#2)} {#3}}}
\nc{\newast}[3] {{\it New\ Astron.\ }{{\bf #1} {(#2)} {#3}}}
\nc{\ncim}[3]{{\it  Nuov.\ Cim.\ }{{\bf #1} {(#2)} {#3}}}
\nc{\np}[3]  {{\it  Nucl.\ Phys.\ }{{\bf #1} {(#2)} {#3}}}
\nc{\pr}[3]  {{\it  Phys.\ Rev.\ }{{\bf #1} {(#2)} {#3}}}
\nc{\pra}[3] {{\it  Phys.\ Rev.\ A\ }{{\bf #1} {(#2)} {#3}}}
\nc{\prb}[3] {{\it  Phys.\ Rev.\ B\ }{{{\bf #1} {(#2)} {#3}}}}
\nc{\prc}[3] {{\it  Phys.\ Rev.\ C\ }{{\bf #1} {(#2)} {#3}}}
\nc{\prd}[3] {{\it  Phys.\ Rev.\ D\ }{{\bf #1} {(#2)} {#3}}}
\nc{\prl}[3] {{\it  Phys.\ Rev.\ Lett.\ }{{\bf #1} {(#2)} {#3}}}
\nc{\pl}[3]  {{\it  Phys.\ Lett.\ }{{\bf #1} {(#2)} {#3}}}
\nc{\prep}[3]{{\it  Phys.\ Rep.\ }{{\bf #1} {(#2)} {#3}}}
\nc{\prsl}[3]{{\it  Proc.\ R.\ Soc.\ London\ }{{\bf #1} {(#2)} {#3}}}
\nc{\ptp}[3] {{\it  Prog.\ Theor.\ Phys.\ }{{\bf #1} {(#2)} {#3}}}
\nc{\ptps}[3]{{\it  Prog\ Theor.\ Phys.\ suppl.\ }{{\bf #1} {(#2)} {#3}}}
\nc{\physa}[3]{{\it Physica\ A\ }{{\bf #1} {(#2)} {#3}}}
\nc{\physb}[3]{{\it Physica\ B\ }{{\bf #1} {(#2)} {#3}}}
\nc{\phys}[3]{{\it  Physica\ }{{\bf #1} {(#2)} {#3}}}
\nc{\rmp}[3] {{\it  Rev.\ Mod.\ Phys.\ }{{\bf #1} {(#2)} {#3}}}
\nc{\rpp}[3] {{\it  Rep.\ Prog.\ Phys.\ }{{\bf #1} {(#2)} {#3}}}
\nc{\sjnp}[3]{{\it  Sov.\ J.\ Nucl.\ Phys.\ }{{\bf #1} {(#2)} {#3}}}
\nc{\sjp}[3] {{\it  Sov.\ J.\ Phys.\ }{{\bf #1} {(#2)} {#3}}}
\nc{\spjetp}[3]{{\it Sov.\ Phys.\ JETP\ }{{\bf #1} {(#2)} {#3}}}
\nc{\yf}[3]  {{\it  Yad.\ Fiz.\ }{{\bf #1} {(#2)} {#3}}}
\nc{\zetp}[3]{{\it  Zh.\ Eksp.\ Teor.\ Fiz.\ }{{\bf #1} {(#2)} {#3}}}
\nc{\zp}[3]  {{\it  Z.\ Phys.\ }{{\bf #1} {(#2)} {#3}}}
\nc{\ibid}[3]{{\sl  ibid.\ }{{\bf #1} {#2} {#3}}}
%
%

\end{document}